# AI-Native 6G for Distributed Intelligence: Traffic Characteristics, Awareness, and AI Grid

Lopamudra Kundu, Xingqin Lin, Shuvo Chowdhury and Sree Sankar

NVIDIA

Email: {lkundu, xingqinl, shuvoc, srsankar}@nvidia.com

*Abstract*— **The sixth-generation (6G) of mobile networks will be shaped not only by artificial intelligence (AI)-enabled network automation and optimization, but also by the need to serve AI as a 6G-native workload. Emerging AI services introduce traffic and compute demands that differ from conventional mobile broadband. Their user experience depends on how quickly useful information is delivered, how bursty and asymmetric multimodal flows are handled, and where inference, retrieval, caching, and content processing are executed. This article presents a joint connectivity-compute view of AI-native 6G. We first characterize representative AI service traffic in terms of uplink/downlink throughput skew, burstiness, and token latency. Next, we discuss how fifth-generation extended reality awareness mechanisms can evolve toward AI traffic characteristics awareness in 6G. Finally, we introduce AI Grid as a distributed AI infrastructure platform for placing workloads according to latency, cost, policy, and service-level constraints. Together, AI-aware connectivity and AI Grid enable 6G as a distributed intelligence platform.**

## I. INTRODUCTION

Artificial intelligence (AI) is rapidly evolving from a cloud-hosted application into a pervasive class of real-time (RT), multimodal, and agentic services [1]. Large language models (LLMs), RT voice assistants, multimodal perception systems, AI agents, code generation tools, and media generation pipelines are increasingly expected to interact with users, devices, enterprise data, and network infrastructures in a continuous and context-aware way. As these services become more deeply embedded into everyday applications, the communications network will no longer carry only conventional, human-generated traffic such as web browsing, video streaming, and/or messaging [2]. Instead, future networks will carry AI service traffic whose structure, timing, and usefulness are tightly coupled to AI model execution [3].

This shift has important implications for sixth-generation (6G) systems. So far, much of the discussion on AI-native 6G has been focused on using AI to improve the network itself [4]. However, an equally important perspective is that 6G must also become native to AI applications. In other words, future 6G systems should be designed not only as networks that use AI, but also as networks that efficiently serve AI [2][5]. This requires understanding of AI workloads both from a connectivity perspective, and a computing perspective.

From the connectivity perspective, some AI applications introduce traffic patterns that differ from traditional mobile broadband services [6]. A text-based chat session may generate relatively small uplink (UL) prompts followed by streamed downlink (DL) tokens. An RT voice assistant may require very low time-to-first-token (TTFT) to preserve conversational quality. An image-generation service may produce highly DL-heavy responses, while video analysis may be dominated by UL media uploads. Agentic AI may trigger chains of model context protocol (MCP) tool calls, retrieval operations, web queries, and intermediate reasoning steps, creating bursty and phase-varying traffic. These examples suggest that AI service traffic is heterogeneous, asymmetric, bursty, and dynamic, and its service quality depends on application-level metrics such as TTFT, time-to-last-token (TTLT), response completeness, and the timely delivery of useful information units (e.g., tokens).

From the computing perspective, AI service quality is also determined by where the workload runs [7]. Network latency, packet loss, and congestion affect user experience, but so do model placement, graphic processing unit (GPU) queueing, cache locality, data residency, cost, and serving capacity. An RT AI assistant may need inference closer to the user to satisfy latency constraints. A vision analytics workload may need to process raw video close to cameras to avoid excessive UL and transport load. A deep reasoning request may benefit from being routed to a larger regional or centralized AI cluster with more capable models and accelerators. Therefore, serving AI-native applications requires a joint connectivity-compute view: the network must know how to carry AI traffic, and the infrastructure must know where to execute AI workloads.

This article develops such an overarching view for AI-native 6G. We first characterize representative AI service traffic across LLM chat, multimodal analysis, agentic AI, and RT conversational AI scenarios. We then discuss how fifth-generation (5G) extended reality (XR) awareness standardized by the third generation partnership project (3GPP) provides a useful starting point for evolving toward AI traffic characteristics awareness in 6G. Next, we extend the discussion from connectivity to computing by introducing AI Grid, a distributed AI infrastructure that can place each workload where it best satisfies latency, cost, policy, and service-level constraints. We also present preliminary AI Grid evaluation results for vision AI workloads, followed by concluding remarks. The central message is that AI-native 6G should become a distributed intelligence platform that jointly optimizes communication and computation. By combining AI traffic characteristics awareness with AI Grid-based workload orchestration, 6G can support a new generation of AI-native applications from day one.

## II. AI SERVICE TRAFFIC CHARACTERISTICS

The traffic in AI-native 6G network is expected to be multimodal, with high degree of heterogeneity across vision (video/image), audio, text, and mixed modality. To evaluate 6G network traffic characteristics generated by various AI services, we look into three classes of AI service scenarios: 1) LLM-based chat and multimodal analysis, 2) agentic AI, and 3) RT conversational AI. Under the first class of AI services, emulated traffic scenarios include basic LLM chat (single-turn), coding chat (LLM generating programming codes), reasoning chat (deep reasoning-based LLM analysis), multimodal analysis (image + PDF as input, and multi-turn text as outputs), image generation (e.g., using DALL-E) and video analysis (content analysis of input video clips). The second class of AI services comprises three types of AI agents with increasing capability of MCP tool calls, viz., shopping agent (maximum 5 tool calls), web search agent (maximum 10 tool calls), and general agent (can access all available MCP tools). The third class of AI services looks into two scenarios of conversational AI, namely RT technical (tech.) assistant (tech. support conversation) and RT audio assistant (voice in/out with text-to-speech simulating voice). Further details are furnished in [8].

| Profile | Delay (ms) | Jitter (ms) | Delay distribution | Loss (%) | Loss distribution | Rate (Mbps) |
|---|---|---|---|---|---|---|
| Ideal | 1 | 0.2 | Normal | 0.001 | Correlated (10%) | 300 |
| Cell center | 20 | 5 | Normal | 0.1 | Correlated (25%) | 100 |
| Cell edge | 120 | 30 | Paretonormal | 1 | Gilbert-Elliot (35%) | 5 |
| Congested | 200 | 50 | Pareto | 3 | Gilbert-Elliot (40%) | 1 |

**Table 1: Network profiles used in the emulation of AI service scenarios [8].**

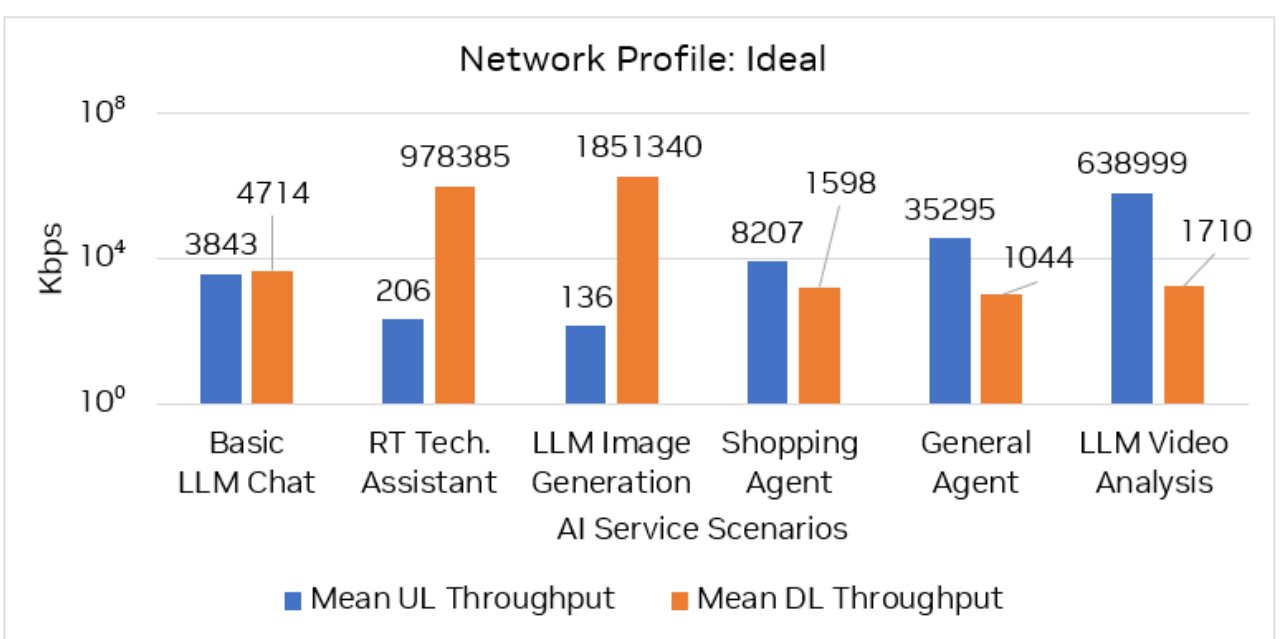

**Figure 1: Variation of AI traffic throughput across scenarios.**

All these AI service scenarios are emulated under a reference network profile called ‘ideal’, with nearly perfect network conditions (i.e., minimum delay/jitter/loss and maximum data rate support), while a subset thereof are further studied under varying network conditions defined by three additional network profiles (‘cell center’, ‘cell edge’ and ‘congested’), with increasingly deteriorating radio propagation conditions. The profile details are provided in Table 1. To quantify network traffic characteristics generated from these AI services, we simulated various scenarios in an open-sourced testbed [8] and collected traffic statistics in terms of three key performance indicators (KPIs) – throughput, burstiness, and token latency. The definition of each of these KPIs, along with the analysis of simulated traffic data are detailed in the following subsections.

### *A. Mean Uplink and Downlink Throughputs*

Mean UL/ DL throughput is the average UL/DL data rate per request, computed as the mean UL/DL payload volume (i.e., request bytes, including user prompts and control metadata in UL, and response bytes including model output and associated metadata in DL), divided by the corresponding request latency (i.e., end-to-end (E2E) application layer response time per AI request, from sending request to receipt of final response), and reported in kilobits per second (kbps) [8].

Figure 1 plots the variation of mean UL and DL throughput (in log scale with base 10) across various AI service scenarios emulated under ideal network profile. Basic LLM chat shows a symmetric UL/DL traffic pattern suggesting nearly equal volume of prompt and response tokens generated in a single-turn chat. The *DL skewness* rapidly grows as the AI service evolves into more complex and interactive scenario like RT tech. assistant, indicating relatively short input prompts and much longer output response streams. Finally, when AI service scenario generates multimodal output (e.g., LLM image generation), the DL skewness grows even more. On the other hand, agentic AI services, e.g., a shopping agent or a general agent, show *UL-heaviness*. LLM video analysis pushes the UL-skewness even higher, since uploading video segments creates much more traffic than text-based model responses.

To summarize, AI service traffic patterns are asymmetric in nature, either being *UL-heavy* or *DL-heavy* depending on the specific scenario. Both the traffic skewness direction (UL/DL) and the degree of skewness can dynamically change within a single AI service session, for example, when a user switches from text-based to multimodal (image/video) chats.

### *B. Peak-to-Mean Burst Ratio and Coefficient of Variation*

Alongside the measurement of average traffic throughput, it is equally crucial to understand the degree of burstiness (i.e., how much the peak traffic overshoots the mean level) and its dispersion (i.e., how the traffic patterns vary over time). To that end, peak-to-mean (P/M) burst ratio, defined as the maximum total per-request traffic volume (request bytes + response bytes) divided by the mean total per-request traffic volume over the measurement interval, indicates the ‘spikiness’ of data burst. Alongside, coefficient of variation (CV), defined as the standard deviation of per-request total traffic volume across requests divided by their mean, indicates the relative dispersion of burstiness over time [8].

Figure 2 top-left subplot shows the variations of burst P/M and burst CV for four AI service scenarios emulated under ideal network profile. Conversational AI, e.g., RT tech. assistant scenario, shows the lowest burstiness, with peak load slightly above average, and CV ≈ 0.2, indicating fairly smooth traffic pattern over time. With agentic AI services, however, the burstiness characteristics change. Shopping agents, for example, show a peak data burst carrying nearly twice the average data volume, and CV ≈ 0.4 indicates more variability in traffic bursts compared to RT tech. assistant. Web search agents make the traffic burstiness even more pronounced. Lastly, the highest degree of burstiness is exhibited by multimodal analysis, with very uneven traffic patterns.

Variation of burstiness with changing network profiles is illustrated for the scenario of web search agent in the top-right subplot of Fig. 2. With deteriorating network conditions, both traffic burstiness and its fluctuations worsen. Near cell edge, for example, higher latency and packet loss cause sharp bursts (P/M ≈ 8) when data finally arrives. The throughput fluctuates widely between quiet periods and high data bursts, rising CV to nearly 1.8. In congested scenario, the burst P/M drops from the extreme cell edge spike but still remains ~ 2x compared to ideal.

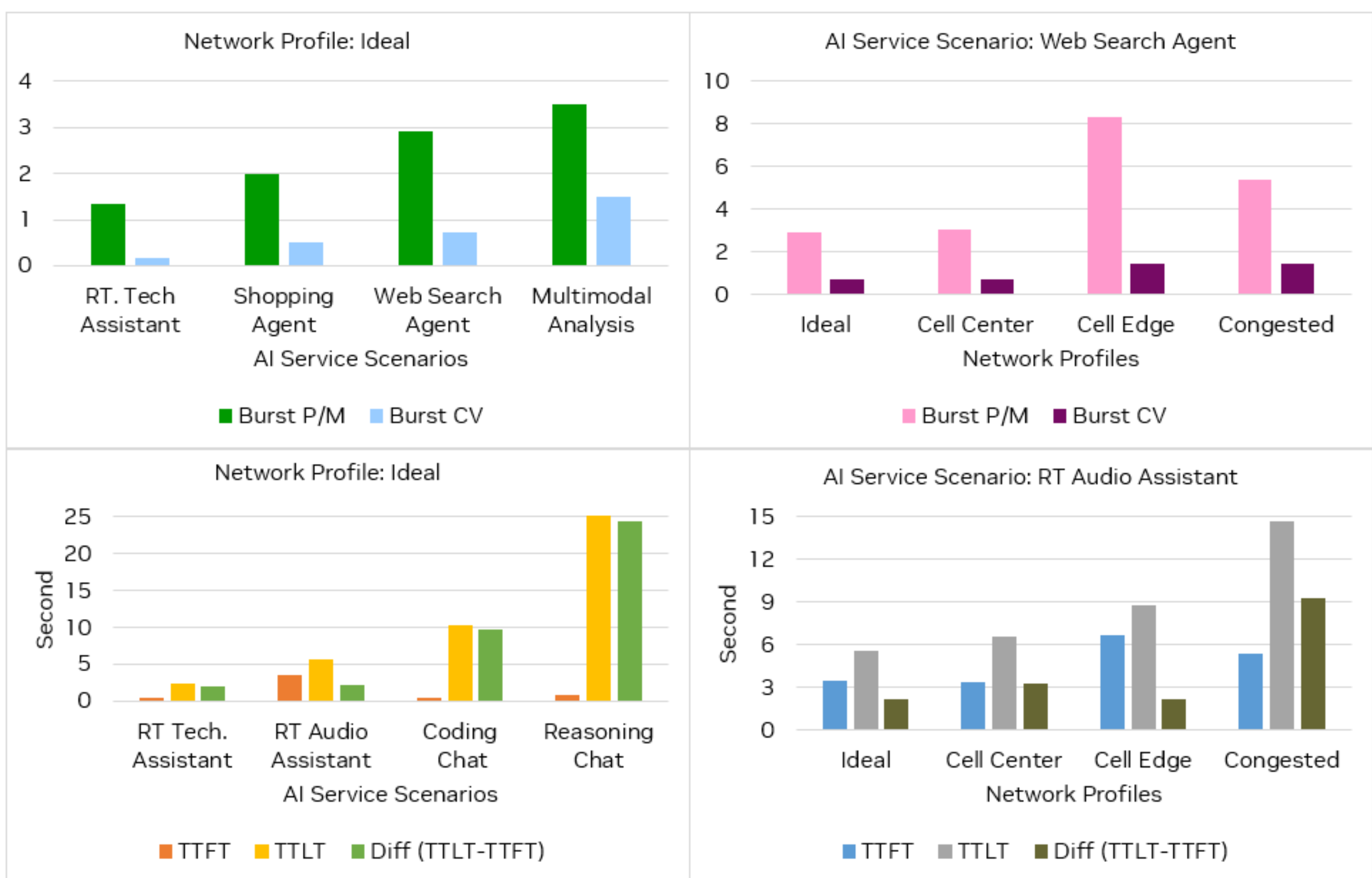

**Figure 2: Variation of AI traffic in terms of token latency and burstiness across various scenarios and network profiles.**

To conclude, degree of burstiness of AI service traffic varies significantly between service types and across network profiles. Depending on the modality, the burstiness may dynamically change within a single AI service session, for example, when switching from text-based chats (e.g., RT tech. assistant) to multimodal analysis, triggering highly bursty traffic patterns.

### *C. Time-to-First Token and Time-to-Last Token Latency*

Quality of experience (QoE) of AI service users not only depends on the volume of traffic generated by a service but also on how responsive the service feels over the network. Token latency becomes critical in that context, determined by three metrics - TTFT, TTLT and their difference. TTFT is the time interval between a user sending a request to the AI service model and receiving the first valid output token from the model, including network delay, queuing, prompt refill and generation time of the first token. TTFT indicates the perceived responsiveness of interactive AI services. TTLT, on the other hand, is the elapsed time between a user sending a request to the AI service model and receiving the final output token of the response, indicating E2E streaming latency for that request. The difference (diff) between TTLT and TTFT signifies the token streaming duration after the reception of the first token, which isolates AI service's token streaming phase from the initial startup latency. All three metrics are measured at 95$^{th}$ percentile (P95), reflecting the tail latency instead of average [8].

Figure 2 depicts the token latency patterns across different AI service scenarios emulated under ideal network profile (bottom-left subplot) as well as across different network profiles (bottom-right subplot) for RT audio assistant, a specific AI service scenario. The lowest TTFT, TTLT and diff are exhibited by RT tech. assistant. Switching to a multimodal AI assistant (i.e., RT audio assistant) notably worsens TTFT to a few seconds, and exacerbates TTLT to >5 seconds while retaining diff at par with RT tech. assistant. For coding-centric chat services, most of the token latency comes from the generation/streaming phase (high TTLT), not from the initial startup (low TTFT). Switching to deep reasoning-based chat elongates TTLT drastically while retaining similar TTFT as coding chat, since complex reasoning requires token generation for a much longer time.

As the network conditions degrade, all three token latency metrics worsen for RT audio assistant. In particular, near cell edge, poorer radio condition delays TTFT with increased RTT, retransmissions and queuing, but the diff is not lengthened commensurately. In fact, TTLT is just a few seconds longer than TTFT, suggesting that the service appears less responsive at the beginning, but once the token streaming starts, it completes within a fairly compact time window. Under network congestion, TTFT improves slightly compared to cell edge, but TTLT goes up significantly, making the diff much larger compared to other network profiles.

In a nutshell, the token latency varies significantly between service types and across network profiles. Depending on the output type/modality, these latencies can dynamically change within a single AI service session, for example, when user transitions from coding-centric chat to deep-reasoning chat, or asks an RT AI assistant to respond via voice instead of text.

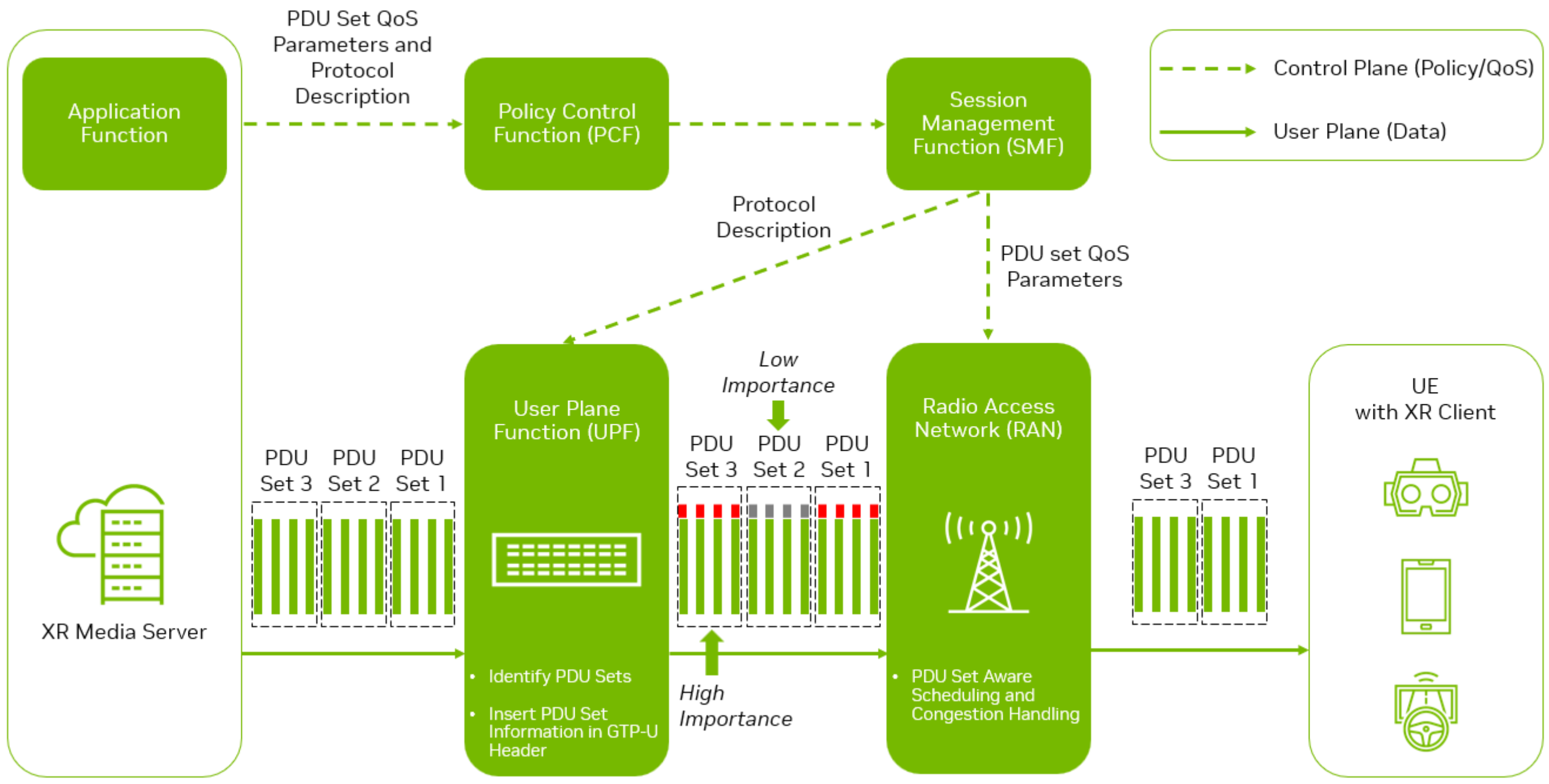

**Figure 3: An illustration of XR awareness with PDU set framework in 5G-Advanced.**

## III. AI SERVICE TRAFFIC CHARACTERISTICS AWARENESS IN 5G-ADVANCED & 6G

The previous section highlighted that AI-native 6G will carry a diverse mix of AI traffics with different latency, throughput, and burstiness profiles. Serving efficiently such diverse AI applications requires mechanisms that let the network be aware of service characteristics, and adapt its resource management accordingly. In this section, we first introduce XR awareness capabilities introduced by 3GPP in 5G-Advanced and then discuss how these mechanisms may evolve to enable AI traffic characteristics awareness in 6G.

### *A. Extended Reality Awareness in 5G-Advanced*

XR performance depends on whether application-relevant units (e.g., frames/slices) arrive on time and with sufficient integrity to be useful. This motivated a tighter coupling between application layer, core network (CN), and radio access network (RAN) [9][10]. A central step in XR awareness is the protocol data unit (PDU) set framework, which standardizes how application-level structure can be exposed to the network. A PDU is the unit of information at a given protocol layer; for example, at the internet protocol (IP) layer, a PDU is an IP packet. 3GPP defines a PDU set as one or more PDUs carrying the payload of one unit of information generated at the application layer (e.g., a video frame or a slice of a frame for XR). Figure 3 illustrates the PDU set concept for XR in 5G-Advanced. An application function (AF) can provide PDU-set assistance information, allowing the user plane function (UPF) in the 5G CN to identify PDU sets and perform marking. The UPF can send PDU set information to the 5G RAN in the general packet radio system (GPRS) tunnelling protocol user plane (GTP-U) header.

In legacy 5G quality-of-service (QoS) framework, the primary abstraction is the QoS flow, identified by a QoS flow indicator (QFI) within a PDU session, and mapped to a 5G QoS indicator (5QI) associating QoS characteristics with packet-level metrics such as packet delay budget (PDB) and packet error rate (PER). That works well when all packets within a QoS flow are equally useful. XR, however, is dominated by application data units (frames/slices) that are fragmented into many packets. Delivering a subset of packets within the time budget may still result in a missed frame deadline or an unusable slice, if subsequent packets of the same frame/slice are delayed. Moreover, not all frames are equally important: losing a reference frame (e.g., an intra-coded (I)-frame) can have a disproportionate impact because it can propagate distortion across subsequent dependent frames (e.g., predictive-coded (P)-frames). With PDU set level QoS handling, the network treats all packets within a PDU set in an integrated manner, and can apply differentiated handling across PDU sets. To this end, 3GPP defined three key PDU set specific QoS parameters: PDU set delay budget (PSDB), PDU set error rate (PSER), and PDU set integrated handling indicator (PSIHI). These parameters are delivered to the RAN through the control plane, analogous to traditional QoS parameters. With (i) PDU set information delivered via the user plane, and (ii) PDU set QoS parameters delivered via the control plane, RAN can apply PDU set aware scheduling and congestion control actions. For example, the network may prioritize allocations that finish a PDU set before PSDB expires, and under congestion may avoid wasting resources on partially delivered sets that are unlikely to be completed in time. The PDU set importance (PSI) field in the PDU set information complements this by enabling graceful degradation: less important PDU sets (e.g., carrying P-frames) can be discarded earlier to preserve resources for more important units (e.g., carrying I-frames).

3GPP complements the PDU set concept with a data burst abstraction: a set of multiple PDUs generated and transmitted in a short time interval. Many XR applications generate traffic in periodic bursts (e.g., one burst per rendered frame interval), and this periodic structure is valuable to the network. In particular, knowledge of burst periodicity and accurate detection of burst boundaries can be exploited for user equipment (UE) power saving.

XR experiences are multimodal in nature. To identify each real-time transport protocol (RTP) media flow among multiplexed, multimodal XR flows, the IP packet filter can include RTP multiplexed media identification information. The AF can provide individual QoS requirements for each media component along with this identification information. In addition, the AF can provide a multimodal service ID (MMSID) together with per-flow QoS information, wherein MMSID explicitly indicates that a set of flows are related, and subject to application coordination.

### *B. Toward AI Traffic Characteristics Awareness in 6G*

AI-native 6G inherits the same fundamental lessons that motivated XR awareness: service experience is determined by application-level structure, not by packet-level behavior alone. However, AI traffic broadens both the diversity and the semantics of that structure. In addition to high-rate, media-like streams, AI services may include, e.g., tokenized interactive inference, multimodal (text/audio/image/video) streams, and sporadic tool-calling/control exchanges. These flows differ not only in burstiness patterns, latency sensitivities and reliability targets, but also in what constitutes usefulness: time-to-first-result versus steady-state throughput, prefix-criticality where early outputs dominate perceived responsiveness, and elasticity where applications can trade precision for timeliness. Efficiently serving this mix requires evolving XR awareness from an XR-specific feature set into a multi-service awareness framework for 6G. However, simply extending XR awareness is not sufficient, as there are fundamental differences in traffic characteristics. XR traffic is predominantly *DL-heavy* and exhibits relatively stable, media-driven patterns, whereas AI traffic can be highly asymmetric and dynamic, with resource demand unpredictably shifting between UL and DL even within a single session. This motivates extending XR awareness into a more adaptive, bidirectional, and semantics-aware framework for AI traffic.

A natural starting point is to generalize the PDU set framework. For XR, a PDU set is a frame or slice; for tokenized inference it can be a token chunk (e.g., a short burst of generated tokens), and for multimodal perception it can be a sensor frame. The core idea is to expose to the network lightweight, relevant AI application semantics so that RAN can prioritize completing useful units. In doing so, the PDU set QoS concepts can be extended as well. The unit-level delay budget naturally maps to AI notions such as TTFT, interactive response deadlines, or maximum allowable timing mismatch among sensor streams used for multimodal fusion. The integrated-handling concept can indicate whether partial delivery remains useful (e.g., partial token chunks may still be valuable, while partial model updates may not), and the importance concept can be used to distinguish critical-path control or early response segments from elastic background traffic.

The second evolution is to broaden data burst awareness into cadence and phase awareness for AI traffic. While XR has shown that periodicity and burst awareness improve scheduling efficiency and UE power saving, AI-service sessions are more variable and less predictable, making deterministic periodicity and burst boundary detection more challenging. As a result, 6G should move beyond reactive signaling toward adaptive and predictive mechanisms. In addition to indicating burst characteristics (e.g., burst size or time-to-next-burst), the network should support detection of higher-level phases, such as latency-critical exchanges, sustained output generation, or background updates, as well as transitions among these phases. User- and application-level profiling can further help anticipate these transitions based on past behavior.

Third, the multiplexed-flow mechanisms introduced for XR should be generalized for modern AI transport and multi-stream sessions. AI services frequently carry multiple logical streams within a single E2E connection. 6G should generalize XR's multiplexed media identification into protocol-aware sub-stream identification, allowing logical AI streams such as prompt upload, token output, audio input, tool call exchange, or background context transfer to be distinguished according to the transport protocol in use. In addition, QoS enforcement should become dynamic at the packet level, allowing QoS requirements to evolve over the lifetime of a stream or sub-stream. Rather than assigning fixed QoS characteristics at session setup, the network should support fine-grained, time-varying QoS signaling that reflects changing application phases. In addition, extending XR's MMSID concept in 6G could give the network a standardized handle to coordinate policy, monitoring, and admission controls across related flows in AI applications.

In summary, XR-aware mechanisms from 5G-advanced may serve as the foundation for designing a unified, semantics-aware framework for supporting AI-native services in 6G. Concepts like PDU sets, set level QoS handling, and burst awareness can be generalized or extended to capture AI-specific characteristics, aiding the network to prioritize what truly determines service usefulness and carry out more efficient and adaptive resource management in 6G.

## IV. AI GRID FOR AI SERVICES

The previous section discussed how AI-native 6G can evolve toward service characteristics-aware connectivity. By exposing lightweight AI application semantics, the network can make better congestion control, resource management, and scheduling decisions. However, connectivity awareness alone is not sufficient. Many AI service bottlenecks are determined not only by how traffic is transported, but also by where the corresponding AI workload is executed. For example, the TTFT of an RT AI assistant not only depends on radio and transport latency, but also on model placement, GPU queueing, cache locality, and serving capacity, among others. Therefore, AI-native 6G requires a computing foundation to complement AI traffic awareness: a distributed AI infrastructure that can place each workload where it best satisfies latency, cost, policy, and service-level constraints/requirements.

### *A. Reference Design*

An AI Grid is a set of geographically distributed and interconnected AI infrastructure nodes that are orchestrated to operate as a unified computing platform [11], as illustrated in Figure 4. Specifically, AI Grid unifies infrastructure tiers including central data centers, regional points of presence (PoPs), telco central offices, mobile switching offices (MSOs), and AI-RAN locations into a programmable fabric for AI-native services. Here, AI-RAN implies the hosting of AI services by RAN infrastructure [12]. This fabric allows AI workloads to move closer to the users, devices, or enterprise premises when locality is critical, while still using larger data centers when scale, model size, or batch efficiency needs outweigh the benefits of Edge proximity. The AI Grid architecture has four pillars: distributed compute, interconnection, control plane, and unified deployment.

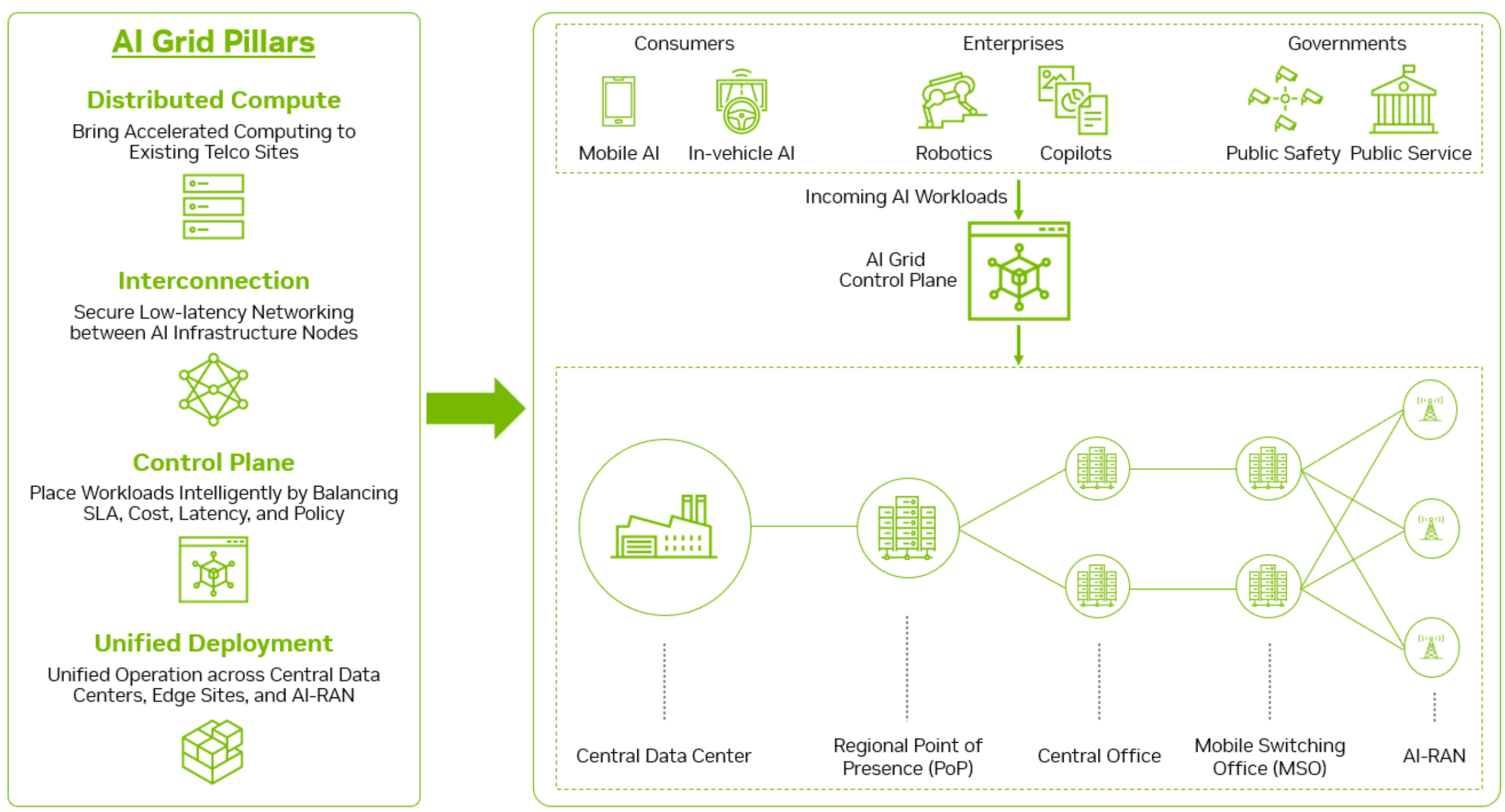

**Figure 4: An illustration of AI Grid reference design for distributed intelligence.**

The first pillar, distributed compute, addresses the geographic mismatch between centralized AI compute and latency- or bandwidth-sensitive AI services. Traditional cloud-centric AI service works well when user requests are tolerant of wide-area network latency, and when input/output traffic volumes are moderate. By deploying accelerated computing across multiple geographically distributed tiers, AI Grid reduces the physical distance between data generation, inference execution, and intelligence consumption nodes. This is especially important for AI workloads that are identified in Section II as UL-heavy, bursty, or latency-sensitive. For instance, video analysis can be placed near the camera or access network to extract metadata locally, while complex reasoning requests can be routed to larger regional or centralized clusters with more capable accelerators.

The second pillar, interconnection, ensures that distributed AI sites do not become isolated Edge silos. AI services frequently require movement of prompts, embeddings, model outputs, and cached states across infrastructure tiers. A high-speed, deterministic-latency wide-area network fabric is therefore needed to connect AI Grid nodes. Note that the user-facing traffic, telco control traffic, storage traffic, and distributed inference traffic have different performance requirements. Separating these traffics helps to ensure that bursty AI workloads do not interfere with time-sensitive telco functions, while still allowing distributed AI execution to scale across multiple GPUs/sites.

The third pillar is AI Grid control plane. In an AI Grid, workload placement and routing are continuously adapted according to requested task, model requirements, latency, cost, cache locality, data governance, and security policy. The AI Grid control plane performs workload-, intent- and resource-aware routing. It evaluates service-level agreements (SLA), latency, cost, model capabilities, and available capacity in real time to determine where each request should be executed. For LLM services, it can also exploit key-value (KV) cache-aware routing by sending cache-compatible prompts to model instances that already hold reusable KV-cache state to reduce token latency and GPU resource consumption. A practical AI Grid control plane can be decomposed into three cooperating functions. First, the AI Grid orchestrator maintains a global view of cluster health, resource quotas, workload placement, and model lifecycle. It provisions models across the grid, detects failures, and enforces desired state under changing conditions. Second, global load balancers along with AI gateway manage ingress traffic, and distribute requests across active instances while considering resource utilization, policy constraints, and SLA requirements. Third, LLM routers add semantic intelligence by analyzing prompt context, complexity, metadata, and potential cache reuse before selecting the most appropriate model or serving location.

The fourth pillar, unified deployment, provides a common operating model across heterogeneous AI infrastructure. A unified AI Grid does not require every site to have identical hardware or software. Instead, each site exposes its capabilities through a common framework so that the AI Grid can place workloads, manage resources, and deliver services in a predictable way. For example, a central data center may serve large models, while regional or Edge sites may provide deterministic QoS to workloads closer to the users and data sources. From the application's perspective, these differences are hidden behind common APIs, policies, and service-level expectations. Thus, unified deployment allows heterogeneous sites to operate as one coordinated AI platform rather than several isolated infrastructure islands.

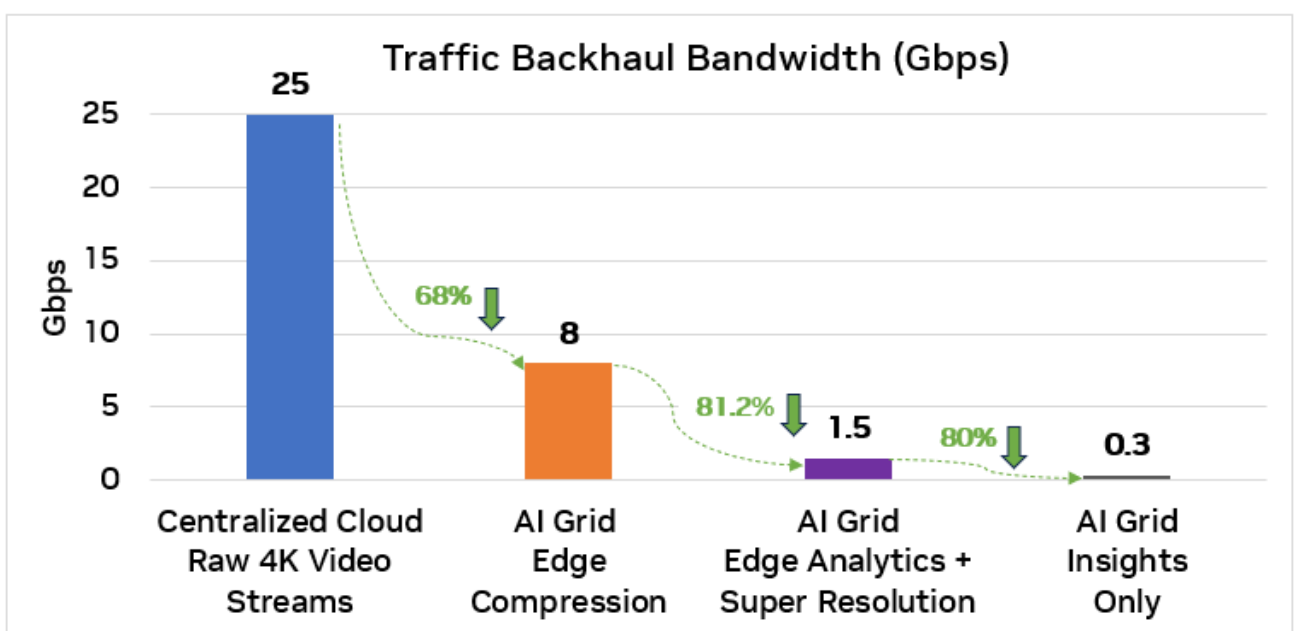

**Figure 5: Backhaul traffic bandwidth variation for vision AI [13].**

In short, NVIDIA's AI Grid reference design extends the AI-native 6G architecture from intelligent connectivity to intelligent workload execution. By combining AI traffic characteristics awareness with AI Grid orchestration, 6G can evolve from a communication-centric network that transports AI traffic into part of a distributed intelligence-centric platform that can jointly optimize connectivity and compute for AI-native applications.

### B. Evaluation Results

To evaluate the efficacy of AI Grid, we consider a vision AI service that aggregates city-scale video feeds, and derives intelligent analytics through inference at the Edge [13]. Such analytics may include public safety information such as detection of unattended objects, human intrusion, and/or traffic incidents. In traditional network infrastructures, aggregating raw video feeds at the central data center generates massive network backhaul costs and inherent delays due to transport of large data volume through multiple hops. Utilizing Edge analytics through distributed intelligence across AI Grid nodes alleviates this problem by keeping raw data localized.

In this section, we present simulation of three distinct AI Grid deployment scenarios [13]. The first one is *Edge compression only*, where compressed videos with sufficient resolution are sent to the Edge for inference. Occasionally, higher resolution frames can be pulled on-demand from the video management systems in such deployments. The second scenario, labeled as *Edge analytics with super resolution*, receives low-resolution video feeds at the Edge, where higher resolution frames are recreated before inference, e.g., by using super-resolution techniques [14], which reduces bandwidth needs compared to the first scenario. The third scenario, called *Insight only*, further distributes intelligence on-prem, which initially processes the raw video frames, and sends primarily metadata (and sometimes, selected frames, if needed) to the Edge for inference, thereby further reducing bandwidth requirements.

Figure 5 shows the backhaul bandwidth impact of running vision AI pipelines with various deployment strategies. The representative deployments consist of 1000 video cameras, each capable of streaming videos with configurable resolutions, and Edge inference running on NVIDIA RTX PRO™ 6000 Blackwell Server Edition GPUs, with the video analytics extracted by NVIDIA Metropolis Blueprint for video search and summarization [15]. While centralized processing of raw video streams can generate significant backhaul traffic volume, moving to Edge compression-based AI Grid reduces the traffic volume considerably, by about 68%. Edge analytics with super resolution can further lower the backhaul traffic load by 81.2% to single-digit gigabits per second (Gbps) range, while switching to primarily metadata instead of always-on video feeds streaming can bring down the bandwidth to only a fraction of Gbps, resulting in an overall 98.8% lower Gbps compared to centralized deployment. Even though these numbers are illustrative, the relative merits between various AI Grid deployment models and central clouds are expected to follow similar trends in emerging real-world deployments.

## V. CONCLUSION AND FUTURE WORK

AI-native 6G should be designed not only as a network that uses AI for automation and optimization, but also as a network that efficiently serves AI-native applications. This article has shown that emerging AI services exhibit highly diverse UL/DL asymmetry, burstiness, and token-latency behavior. These characteristics are not static; they can change within a single session as the application transitions between phases and/or modalities. Building on the service-awareness mechanisms introduced for XR in 5G-Advanced, 6G can evolve toward AI traffic characteristics awareness. By exposing lightweight application semantics, the network can prioritize traffic according to what actually determines AI service quality. However, AI service experience is also strongly shaped by where computation is performed. AI Grid complements AI-aware connectivity by providing a distributed, interconnected, orchestrated, and unified AI infrastructure platform that intelligently places each workload where it performs the best.

Future works should advance this connectivity-compute convergence in several directions. First, broader AI traffic measurement campaigns are needed across model families, modalities, device types, and deployment conditions. Second, standardizable abstractions are required for exposing AI service semantics to the network without leaking sensitive application or user information. Third, joint optimization algorithms should be developed for coordinating radio scheduling, edge inference placement, and cache-aware serving. Finally, E2E prototypes that integrate AI-aware RAN/CN mechanisms with AI Grid orchestration will be useful for validating AI-native 6G as a distributed intelligence platform.

## BIOGRAPHIES

Lopamudra Kundu (lkundu@nvidia.com) is a Senior Standards Engineer at NVIDIA, engaged in 3GPP 5G/6G standardization. Her research areas intersect AI, accelerated computing, digital signal processing and RAN.

Xingqin Lin (xingqinl@nvidia.com) is a Senior 3GPP Standards Engineer at NVIDIA, where he focuses on 3GPP standardization and conducts research at the intersection of 5G/6G and AI.

Shuvo Chowdhury (shuvoc@nvidia.com) is a Principal Product Manager at NVIDIA, responsible for building AI-RAN & Edge AI platforms integrating AI, 5G/6G and Cloud at NVIDIA.

Sree Sankar (srsankar@nvidia.com) is a Global Head for AI Grid at NVIDIA, where she leads the strategy and vision for NVIDIA's distributed inference platform.